\documentclass[]{agujournal}
\journalname{Geophysical Research Letters}

\usepackage{verbatim}

\begin{document}

%% ------------------------------------------------------------------------ %%
%  Title
% 
% (A title should be specific, informative, and brief. Use
% abbreviations only if they are defined in the abstract. Titles that
% start with general keywords then specific terms are optimized in
% searches)
%
%% ------------------------------------------------------------------------ %%

% Example: \title{This is a test title}

\title{Exact Vlasov-Maxwell equilibria for asymmetric current sheets}

%% ------------------------------------------------------------------------ %%
%
%  AUTHORS AND AFFILIATIONS
%
%% ------------------------------------------------------------------------ %%

% Authors are individuals who have significantly contributed to the
% research and preparation of the article. Group authors are allowed, if
% each author in the group is separately identified in an appendix.)

% List authors by first name or initial followed by last name and
% separated by commas. Use \affil{} to number affiliations, and
% \thanks{} for author notes.  
% Additional author notes should be indicated with \thanks{} (for
% example, for current addresses). 

% Example: \authors{A. B. Author\affil{1}\thanks{Current address, Antartica}, B. C. Author\affil{2,3}, and D. E.
% Author\affil{3,4}\thanks{Also funded by Monsanto.}}

\authors{O. Allanson\affil{1}\affil{2}, F. Wilson\affil{1}, T. Neukirch\affil{1}, Y.-H. Liu\affil{3} and J.D.B. Hodgson\affil{1}}

% \affiliation{1}{First Affiliation}
% \affiliation{2}{Second Affiliation}
% \affiliation{3}{Third Affiliation}
% \affiliation{4}{Fourth Affiliation}

\affiliation{1}{Solar \& Magnetospheric Theory Group, School of Mathematics \& Statistics, University of St Andrews, St Andrews, KY16 9SS, UK}
\affiliation{2}{Space and Atmospheric Electricity Group, Department of Meteorology, University of Reading, Reading, RG6 6BB, UK}
\affiliation{3}{NASA-Goddard Space Flight Center, Greenbelt, Maryland 20771, USA}

%(repeat as many times as is necessary)

%% Corresponding Author:
% Corresponding author mailing address and e-mail address:

% (include name and email addresses of the corresponding author.  More
% than one corresponding author is allowed in this LaTeX file and for
% publication; but only one corresponding author is allowed in our
% editorial system.)  

% Example: \correspondingauthor{First and Last Name}{email@address.edu}

\correspondingauthor{Oliver Allanson}{o.allanson@reading.ac.uk}

%% Keypoints, final entry on title page.

% Example: 
% \begin{keypoints}
% \item	List up to three key points (at least one is required)
% \item	Key Points summarize the main points and conclusions of the article
% \item	Each must be 100 characters or less with no special characters or punctuation 
% \end{keypoints}

%  List up to three key points (at least one is required)
%  Key Points summarize the main points and conclusions of the article
%  Each must be 100 characters or less with no special characters or punctuation 

\begin{keypoints}
\item New exact Vlasov-Maxwell equilibria for 1D asymmetric current sheets.
\item Written as a combination of four shifted Maxwellian distribution functions.
\item Self-consistent with asymmetric magnetic field, temperature, density and pressure profiles.
\end{keypoints}

%% ------------------------------------------------------------------------ %%
%
%  ABSTRACT
%
% A good abstract will begin with a short description of the problem
% being addressed, briefly describe the new data or analyses, then
% briefly states the main conclusion(s) and how they are supported and
% uncertainties. 
%% ------------------------------------------------------------------------ %%

%% \begin{abstract} starts the second page 

\begin{abstract}
The NASA Magnetospheric Multiscale mission has made in-situ diffusion region and kinetic-scale resolution measurements of asymmetric magnetic reconnection for the first time \citep{Burch-2016Science}, in the Earth's magnetopause. The principal theoretical tool currently used to model collisionless asymmetric reconnection is particle-in-cell simulations. Many particle-in-cell simulations of asymmetric collisionless reconnection start from an asymmetric Harris-type magnetic field, but with distribution functions that are not exact equilibrium solutions of the Vlasov equation. We present new and exact equilibrium solutions of the Vlasov-Maxwell system that are self-consistent with one-dimensional asymmetric current sheets, with an asymmetric Harris-type magnetic field profile, plus a constant non-zero guide field. The distribution functions can be represented as a combination of four shifted Maxwellian distribution functions. This equilibrium describes a magnetic field configuration with more freedom than the previously known exact solution \citep{Alpers-1969}, and has different bulk flow properties.
\end{abstract}

%% ------------------------------------------------------------------------ %%
%
%  TEXT
%
%% ------------------------------------------------------------------------ %%

%%% Suggested section heads:
% \section{Introduction}
% 
% The main text should start with an introduction. Except for short
% manuscripts (such as comments and replies), the text should be divided
% into sections, each with its own heading. 

% Headings should be sentence fragments and do not begin with a
% lowercase letter or number. Examples of good headings are:

% \section{Materials and Methods}
% Here is text on Materials and Methods.
%
% \subsection{A descriptive heading about methods}
% More about Methods.
% 
% \section{Data} (Or section title might be a descriptive heading about data)
% 
% \section{Results} (Or section title might be a descriptive heading about the
% results)
% 
% \section{Conclusions}

\section{Introduction}
The formation of current sheets is ubiquitous in plasmas. These current sheets form between plasmas of different origins that encounter each other, such as at Earth's magnetopause between the magnetosheath and magnetospheric plasmas \citep{Dungey-1961, Phan-1996}; or they develop spontaneously in magnetic fields that are subjected to random external drivings \citep{Parker-1994}, such as in the solar corona region. Under most circumstances, the plasma conditions on either side of the current sheet can be different, e.g. the magnetic field strength and orientation. Such current sheets are dubbed asymmetric. Asymmetric current sheets are also observed at Earth's magnetotail \citep{Oieroset-2004}, in the solar wind \citep{Gosling-2006}, between solar flux tubes \citep{Linton-2006, Murphy-2012, Zhu-2015}, in turbulent plasmas \citep{Servidio-2009, Karimabadi-2013}, and inside tokamaks \citep{Kadomtsev-1975}. 

As per Poynting's theorem \citep{Poynting-1884, Birn-2010}, these intense current sheets are ideal locations for magnetic energy conversion and dissipation \citep{Zenitani-2011}. The dominant mechanisms that release the free energy include magnetic reconnection, and various plasma instabilities. The asymmetric feature has now been included in modelling the reconnection rate \citep{Cassak-2007}, the development of the lower-hybrid instability \citep{Roytershteyn-2012} and the suppression of reconnection at Earth's magnetopause \citep{Swisdak-2003, Phan-2013, Trenchi-2015, Liu-2016}. The physics in the linear stage could affect the dynamical evolution of the current sheets \citep{Dargent-2016}. Thus, developing an exact Vlasov equilibrium for the current sheet is important, but it is challenging. The well-known solution of the symmetric Harris sheet \citep{Harris-1962} has been extended to the relativistic regime \citep{Hoh-1966}, the Kappa distribution \citep{Fu-2005}, and later the force-free limit \citep{Harrison-2009PRL, Wilson-2011, Stark-2012, Abraham-Shrauner-2013, Allanson-2015POP, Kolotkov-2015, Allanson-2016JPP}. In this letter, we present a new exact Vlasov-Maxwell equilibrium solution for asymmetric current sheets.

The intention of the exact solution that we present in this paper is to represent a step forward in the analytical modelling of asymmetric Vlasov-Maxwell equilibria, that is of particular relevance to particle-in-cell (PIC) simulations and analysis using kinetic theory. Inevitably, working within the confines of an exact model does imply that we cannot accurately represent all desired features of the magnetopause current sheet system, and some of these restrictions will be discussed.

\subsection{The current sheet equilibrium}\label{sec:equilibrium}
The specific magnetic field profile that we consider is a one-dimensional (1D) current sheet, composed of an `asymmetric Harris sheet' with a constant guide field, such as that first used in analytical study of the tearing mode at the dayside Magnetopause in \citet{Quest-1981B}. In \emph{mks} units and $(\hat{x},\hat{y},\hat{z}) \sim (\hat{L},\hat{M},\hat{N})$ coordinates (e.g. see \citet{Hapgood-1992}), the vector potential, magnetic field and current density for the `asymmetric Harris sheet plus guide' (AH+G) model can be written
\begin{eqnarray}
\boldsymbol{A}(\tilde{z})&=&B_0L(\,C_3\tilde{z},\, -C_1\tilde{z}-C_2\ln \cosh\tilde{z},\,0),\nonumber\\
\nabla\times\boldsymbol{A}=\boldsymbol{B}(\tilde{z})&=&B_0(\,C_1+C_2\text{tanh}\tilde{z},\,C_3,\,0 ),\label{eq:magfield}\\
\frac{1}{\mu_0}\nabla\times\boldsymbol{B}=\boldsymbol{j}(\tilde{z})&=&\frac{B_0}{\mu_0L}(\, 0,\, C_2\text{sech}^2\tilde{z},\, 0),\label{eq:current}
\end{eqnarray}
respectively, with $\mu_0$ the magnetic permeability \emph{in vacuo}; $C_1, C_2$ and $C_3\ne 0$ dimensionless constants; and $B_0$ and $L$ dimensional constants that normalise the vector potential $(\boldsymbol{A}=B_0L\tilde{\boldsymbol{A}})$, magnetic field $(\boldsymbol{B}=B_0\tilde{\boldsymbol{B}})$, current density $(\boldsymbol{j}=j_0\tilde{\boldsymbol{j}})$ and $z$ $(z=L\tilde{z})$, with $j_0=B_0/(\mu_0L)$. 

\begin{figure}[h!]
 \centering
% when using pdflatex, use pdf file:
 \includegraphics[width=15pc]{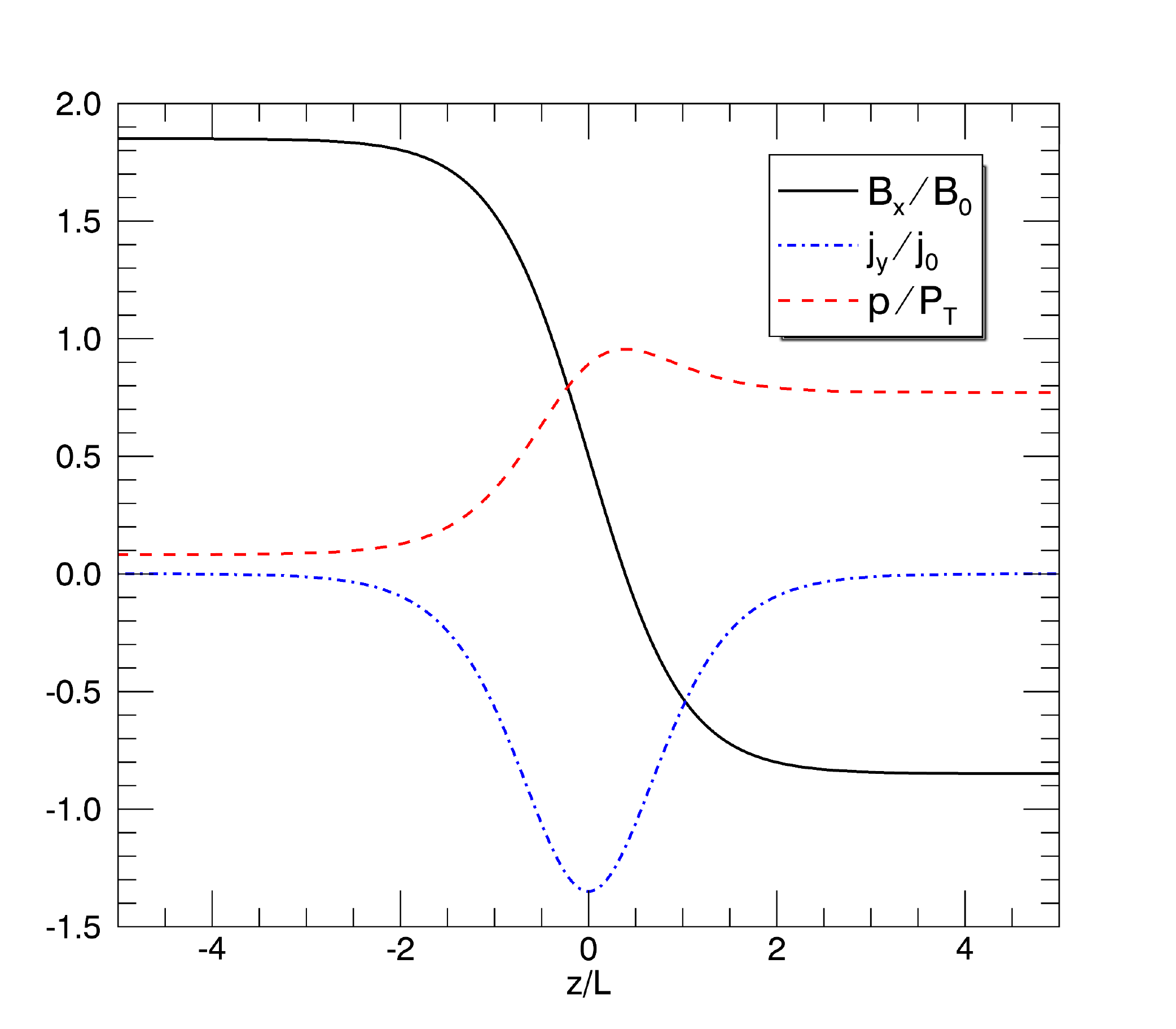}
%
% when using dvips, use .eps file:
% \includegraphics[width=20pc]{figsamp.eps}
%
 \caption{Normalised magnetic field $\tilde{B}_x$, current density $\tilde{j}_y$, and scalar pressure $\tilde{p}$ for \emph{Parameter Set One}.  }
 \label{figone}
  \end{figure}

\begin{figure}[h!]
 \centering
% when using pdflatex, use pdf file:
 \includegraphics[width=15pc]{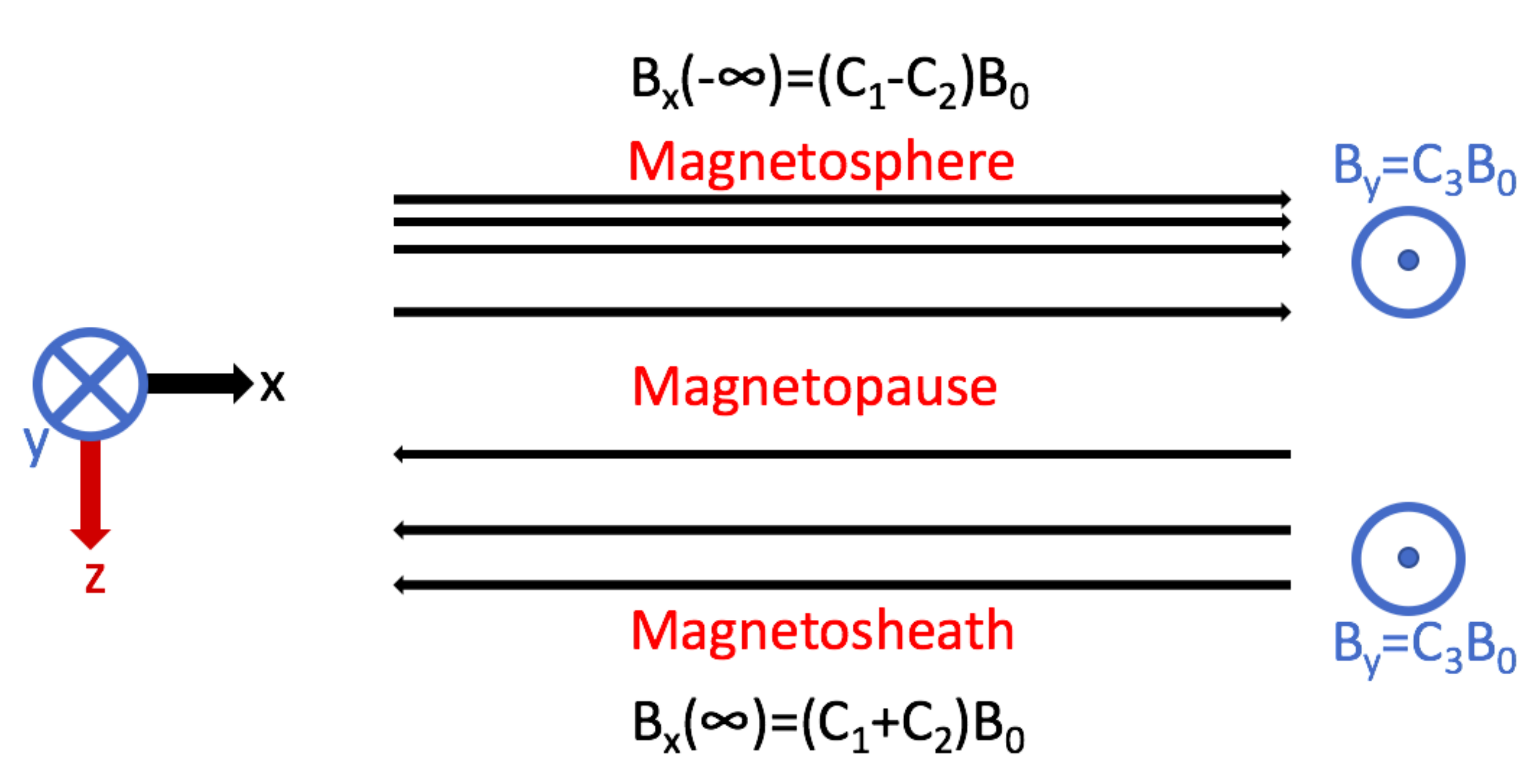}
%
% when using dvips, use .eps file:
% \includegraphics[width=20pc]{figsamp.eps}
%
 \caption{A representative diagram of the equilibrium magnetic field, for $C_{1}+C_{2}<0$, $C_{1}-C_{2}>0$ and $C_3<0$. }
 \label{figtwo}
  \end{figure}

The fluid equilibrium for the AH+G current sheet is maintained by the gradient of a scalar pressure, $p=p(z)$, according to $\nabla p = \boldsymbol{j}\times\boldsymbol{B}$ and $d/dz[p+B^2/(2\mu_0)]=0$. The scalar pressure in force balance with the AH+G field is given by
\begin{equation}
p(\tilde{z})=P_{T}-\frac{B_0^2}{2\mu_0}\left(C_1^2+2C_1C_2\tanh \tilde{z}+C_2^2\tanh^2\tilde{z}+C_3^2\right),\label{eq:scalar_pressure}
\end{equation}
for $P_{T}$ the total pressure (magnetic plus thermal), and $p(z)>0$ for $C_1^2+2|C_1C_2|+C_2^2+C_3^2<2\mu_0P_{T}/B_0^2$. Example profiles of $\tilde{B}_x$, $\tilde{j}_y$ and $\tilde{p}(\tilde{z})=p/P_T$ are plotted in Figure \ref{figone} for parameter values $C_1=0.5$, $C_2=-1.35$, $C_3\approx -0.42$, and $P_T\approx 3.92B_0^2/(2\mu_0)$, and hereafter referred to as \emph{Parameter Set One}. For \emph{Parameter Set One}, the left and right hand sides of the plot could represent the magnetosphere and magnetosheath respectively, whilst the central current sheet is in the magnetopause (see Figure \ref{figtwo} for a representative diagram of the equilibrium configuration). \emph{Parameter Set One} corresponds to magnetic field asymmetry, total magnetic shear, and number density/scalar pressure asymmetries of 
\begin{eqnarray}
B_{\text{ratio}}=\frac{|\boldsymbol{B}_{\text{sphere}}|}{|\boldsymbol{B}_{\text{sheath}}|}=2,\hspace{3mm}\phi_{B,\text{shear}}=\cos^{-1}(\hat{\boldsymbol{b}}_{\text{sphere}}\cdot\hat{\boldsymbol{b}}_{\text{sheath}})\approx 140^{\circ}\nonumber\\
n_{\text{ratio}}=\frac{n_{\text{sheath}}}{n_{\text{sphere}}}=p_{\text{ratio}}=\frac{p_{\text{sheath}}}{p_{\text{sphere}}}\approx9.50 , \nonumber
\end{eqnarray}
with $\hat{\boldsymbol{b}}$ the magnetic field unit vector, the \emph{sheath}/\emph{sphere} subscripts denoting $z=\infty,-\infty$ respectively. These asymmetries show positive similarities with certain magnetopause properties, given typical magnetopause conditions (e.g. see \citet{Burch-2016Science, Hesse-2016}). We stress that these asymmetries relate to a particular selction of parameters, which are chosen to demonstrate an example of the types of asymmetric conditions that the distribution function (DF) can support.

The ratio of the number densities was derived using a relation, $p(\tilde{z})=C n(\tilde{z})$, for $C$ a constant. This `fluid' relation is valid even for the Vlasov model that we shall derive, but this does not mean that the `kinetic temperature' is constant, and merits the following discussion. The macroscopic force balance self-consistent with a quasineutral Vlasov equilibrium is maintained by the divergence of a rank-2 pressure tensor, $P_{ij}=P_{ij}(A_x(z),A_y(z))$ (e.g. see \citet{Channell-1976, Mynick-1979a, Schindlerbook}), according to $\nabla\cdot\boldsymbol{P}=\boldsymbol{j}\times\boldsymbol{B}$. Hence, $p=nk_BT$ is in principle an approximation to the kinetic physics, with the pressure and temperature properly defined by rank-2 pressure tensors. However, in our geometry, the scalar pressure that maintains fluid equilibrium is identified with the pressure tensor component that is self-consistent with a kinetic equilibrium, according to $p=P_{zz}$ (e.g. see \cite{Harrison-2009PRL}), giving 
\begin{equation}
\frac{d}{dz}\left(P_{zz}+\frac{B^2}{2\mu_0} \right)=0.  \label{eq:pressure_balance}
\end{equation}
Note that $P_{zz}$ is not the only non-zero component of $P_{ij}$, but it is the only component that plays a role in the force-balance of the equilibrium. It can be shown \citep{Channell-1976} that for 1D Vlasov-Maxwell equilibria like that considered in this paper, $p=P_{zz}=Cn$ holds, and so our expression for $n_{\text{ratio}}$ is correct for both the fluid and kinetic approaches. In Section \ref{sec:plots} we shall use other components of $P_{ij}$ to define the kinetic temperature, which is asymmetric, as plotted in Figure \ref{figfive}.

The AH+G magnetic field is very similar to a magnetic field introduced in the Appendix of \citet{Alpers-1969}, in a rotated coordinate system: the AH+G field defined in equation (\ref{eq:magfield}) reproduces the `Alpers magnetic field' under a rotation $\tan \theta =C_1/C_3$. However, the Alpers magnetic field has one fewer degree of freedom (i.e. an extra constraint on $C_1, C_2, C_3$).

\subsection{Non-equilibrium initial conditions for PIC simulations} \label{sec:noneq}
In the effort to model asymmetric magnetopause reconnection, fields such as the Alpers and AH+G models, and variations that could involve a `double' current sheet structure and/or no guide field have been used in PIC simulations in e.g. \citet{Swisdak-2003, Pritchett-2008, Huang-2008, Malakit-2010, Wang-2013, Aunai-2013, Hesse-2013, Hesse-2014, Dargent-2016, Liu-2016}. All of these studies except that of \cite{Dargent-2016} have used `flow-shifted' Maxwellian DFs as initial conditions
\begin{equation}
f_{\text{Maxw},s}(z,\boldsymbol{v})=\frac{n(z)}{(\sqrt{2\pi}v_{th,s})^3}\exp\left[\frac{\left(\boldsymbol{v}-\boldsymbol{V}_s(z)\right)^2}{2v_{th,s}^2}\right], \label{eq:Maxshift}
\end{equation}   
with $v_{th,s}$ a characteristic value of the thermal velocity of species $s$, $\boldsymbol{V}_{s}$ the bulk velocity of species $s$, and $n(z)$ a number density. These DFs can reproduce the same moments $(n(z), \boldsymbol{V}_s(z), p(z))$ necessary for a quasineutral fluid equilibrium.

Despite the fact that the DF, $f_{\text{Maxw},s}$,  in equation (\ref{eq:Maxshift}) reproduces the desired moments, it is not an exact solution of the Vlasov equation and hence does not describe a kinetic equilibrium. As explained in \citet{Aunai-2013} on the subject of particle-in-cell (PIC) simulations, the fluid equilibrium characterised by a flow-shifted Maxwellian can evolve to a quasi-steady state \emph{``with an internal structure very different from the prescribed one''}, and as demonstrated in \citet{Pritchett-2008}, undesired electric fields, $E_z$, \emph{``coherent bulk oscillations''} and other perturbations may form. 

The main aim of this paper is to calculate exact solutions of the equilibrium Vlasov-Mawell equations consistent with the AH+G magnetic field in equation (\ref{eq:magfield}), in order to circumvent the need to use non-kinetic-equilibrium DFs of the form in equation (\ref{eq:Maxshift}) as initial conditions in collisionless PIC simulations of asymmetric reconnection.

\subsection{Two prior Vlasov-Maxwell equilibria for asymmetric current sheets }
In the Appendix to \citet{Alpers-1969}, a DF is derived that is consistent with the Alpers magnetic field (as described in Section \ref{sec:equilibrium}). As is necessary for consistency between the microscopic and macroscopic descriptions, the Alpers DF is self-consistent with the prescribed magnetic field, i.e. the sum of the individual species (kinetic) currents are equal to the current prescribed by Amp\`{e}re's Law, i.e. $\sum_s \boldsymbol{j}_{s}=\boldsymbol{j}=\nabla\times\boldsymbol{B}/\mu_{0}$. However, the $\boldsymbol{j}_{s}$ are non-zero at $z=+\infty$ (in our co-ordinates), i.e. the magnetosheath side. In contrast, equation (\ref{eq:current}) shows that the macroscopic current densities vanish as $z\to\pm\infty$, i.e. the Alpers DF gives species currents $\boldsymbol{j}_{s}$ that are not proportional to the macroscopic current $\boldsymbol{j}$. That is to say that there is finite ion and electron mass flow at infinity. This could be appropriate if one wishes to consider a larger scale/global magnetopause model that includes flows at the boundary corresponding to the magnetosheath, for example, but it might not be appropriate if one wishes to consider the domain as a `patch', representing a current sheet structure locally (whilst formally speaking, the spatial domain in our model is infinite, this is not necessarily intended to reproduce the entire spatial domain of the solar wind-magnetosheath-magnetopause-magnetosphere system). The non-vanishing of the individual species bulk flows at the boundaries in the Alpers equilibrium are also inconsistent with most of the initial conditions of typical PIC simulations of asymmetric reconnection, \emph{viz.}, in the absence of an exact Vlasov equilibrium the simulations are typically initiated with a shifted-Maxwellian consistent with zero species flow at the boundary. The DF that we derive shall be consistent macroscopically with an equilibrium for which there are no mass flows at infinity, and is self-consistent with a magnetic field that has more degrees of freedom than that in \citet{Alpers-1969}. 

The second relevant work is that of \cite{Belmont-2012}, in which `semi-analytic' Vlasov-Maxwell equilibria are found numerically. The magnetic field in that paper is actually a symmetric Harris sheet without guide field, i.e. $C_1=C_3=0$, but with asymmetric profiles of the density, pressure and temperature. The DFs calculated therein are not found using a typical constants of motion approach as is to be used in this paper. Instead, they are found by considering ion DFs, such that when expressed in terms of the motion invariants, are double-valued functions. The `semi-analytic' DF that is derived by \cite{Belmont-2012} has been used as the initial condition for PIC simulations in \cite{Dargent-2016}. The model was generalised by \citet{Dorville-2015} to include a magnetic field profile similar to the force-free Harris sheet \citep{Harrison-2009PRL}, and also an electric field profile.

\section{New Vlasov-Maxwell equilibrium for asymmetric current sheets}
\subsection{Channell's method}
The AH+G equilibrium defined by equations (\ref{eq:magfield}) and (\ref{eq:pressure_balance}) is translationally invariant in the $xy$ plane, giving rise to two conserved canonical momenta for particles of species $s$, $p_{xs}=m_sv_{x}+q_sA_{x}$, $p_{ys}=m_sv_y+q_sA_y$. Because we are considering an equilibrium, the particle Hamiltonian of species $s$ is also conserved, $H_s=m_sv^2/2+q_s\phi$, for $\phi$ the electrostatic potential. Jeans' theorem implies that one can always solve the Vlasov equation by choosing $f_s$ to be a function of known constants of motion (\citet{Jeans-1915, Lynden-Bell-1962}), and the solution will be physically meaningful provided $f_s\ge 0$ and velocity-space moments of all order exist (\citet{Schindlerbook}). Using this fact, and assumptions common to much theoretical work on one-dimensional ($1D$) translationally invariant Vlasov-Maxwell equilibria (e.g. see \citet{Alpers-1969, Channell-1976, Schindlerbook, Harrison-2009PRL, Wilson-2011, Abraham-Shrauner-2013, Kolotkov-2015, Allanson-2015POP, Allanson-2016JPP}), we assume $\phi=0$ (`strict neutrality'), and that
\begin{equation}
f_s(H_{s},p_{xs},p_{ys})=\frac{n_{0s}}{(\sqrt{2\pi}v_{th,s})^3}e^{-\beta_sH_s}g_s(p_{xs},p_{ys}), \label{eq:ansatz}
\end{equation}
for $n_{0s}$ a constant with dimensions of number density, $\beta_s=1/(m_sv_{th,s}^2)$, $m_s$ the mass and $g_s$ an unknown function of the canonical momenta for particle species $s$, which is yet to be determined. Calculating self consistent $g_s$ functions (and hence Vlasov equilibrium DFs) for a given macroscopic equilibrium is an example of the `inverse problem in collisionless equilibria' (e.g. see \citet{Channell-1976,Allanson-2016JPP}), for which there is not necessarily a guaranteed exact solution. The method that we shall use is known as `Channell's method' \citep{Channell-1976} which is used in many of the works listed above, and has been somewhat generalised in \citet{Mottez-2003}. We note that a treatment of this inverse problem is given in \citet{Alpers-1969} that is very similar to that of Channell. The major benefit of using Channell's method for this problem is that we obtain an exact solution that is readily implementable, but one downside is that the asymmetry of the number density is directly tied to that of the magnetic field, i.e. there can be no asymmetry in the density profile when $C_1=0$. This is in contrast to the numerical methods used by \citet{Belmont-2012, Dorville-2015}.

The method rests on calculating a functional form of $P_{zz}(A_x,A_y)$ that `reproduces' the scalar pressure of equation (\ref{eq:scalar_pressure}) as a function of $z$, i.e. $P_{zz}(A_x,A_y)(z)=p(z)$, but also that satisfies $\partial P_{zz}/\partial \boldsymbol{A}=\boldsymbol{j}(z)$ (for fuller details on the background theory of this first and crucial step, see e.g. \citet{Mynick-1979a,Schindlerbook, Harrison-2009POP}). There could in principle be infinitely many functions $P_{zz}(A_x,A_y)$ that
satisfy both the criteria necessary for Channell's method, however we shall choose a specific $P_{zz}(A_x,A_y)$ which allows us to make analytical progress. 

Similar to the procedure in \citet{Alpers-1969}, by substituting linear combinations of two distinct representations of $\tanh\tilde{z}(A_x,A_y)$,
\begin{eqnarray}
\tanh\tilde{z}&=&1-e^{-\tilde{z}}\text{sech}\tilde{z}=1-e^{\frac{C_1-C_2}{C_2C_3}\tilde{A}_x}e^{\frac{1}{C_2}\tilde{A}_y},\nonumber\\
\tanh\tilde{z}&=&\sqrt{1-\text{sech}^2\tilde{z}}=\sqrt{1-e^{\frac{2C_1}{C_2C_3}\tilde{A}_x}e^{\frac{2}{C_2}\tilde{A}_y}},\nonumber
\end{eqnarray}
into equation (\ref{eq:scalar_pressure}), we arrive at
\begin{eqnarray}
P_{zz}(\tilde{A}_x,\tilde{A}_y)&=&P_{T}-\frac{B_0^2}{2\mu_0}\Bigg\{    C_1^2+C_3^2+2C_1C_2\left(  1-e^{\frac{C_1-C_2}{C_2C_3}\tilde{A}_x}e^{\frac{1}{C_2}\tilde{A}_y}    \right)   \nonumber\\
&&+C_2^2\left[ k\left( 1-e^{\frac{C_1-C_2}{C_2C_3}\tilde{A}_x}e^{\frac{1}{C_2}\tilde{A}_y}   \right)^2+(1-k)\left(  1-e^{\frac{2C_1}{C_2C_3}\tilde{A}_x}e^{\frac{2}{C_2}\tilde{A}_y} \right)      \right]    \Bigg\},\label{eq:pressure_tensor}
\end{eqnarray}
for $k$ a constant. This form of $P_{zz}$ satisfies $\partial P_{zz}/\partial A_x(\tilde{z})=0$ and $\partial P_{zz}/\partial A_y(\tilde{z})=B_0C_2/(\mu_0L)\text{sech}^2\tilde{z}$ when $k=C_1/C_2$, and is positive over all $(A_x,A_y)$ when $C_{1}C_{2}<0$ and $(C_1-C_{2})^2+C_3^2<2\mu_0P_{T}/B_0^2$. 

Next we use the assumed form of the DF in equation (\ref{eq:ansatz}) in the definition of the pressure tensor component $P_{zz}$ as the second-order velocity moment of the DF, $P_{zz}=\sum_sm_s\int v_z^2f_sd^3v$. Note that the pressure tensor should be written as the second order moment of $f_s$ by $\boldsymbol{w}_s^2=\left(\boldsymbol{v}-\boldsymbol{V}_s\right)^2$, but the DF (equation (\ref{eq:ansatz})) is an even function of $v_z$, which implies that $V_{zs}=0$. When the dependence of $f_s$ on the Hamiltonian, $H_s$, is given by $\exp(-\beta_sH_s)$ as it is here, the integral equation for $P_{zz}$ can be interpreted \citep{Allanson-2016JPP} as a Weierstrass transform (e.g. see \citet{Bilodeau-1962}), and can be amenable to solution by Fourier transforms (e.g. see \cite{Harrison-2009PRL,Abraham-Shrauner-2013}), or expansion of $g_s$ in Hermite polynomials (e.g. see \citet{Abraham-Shrauner-1968,Hewett-1976,Channell-1976,Suzuki-2008,Allanson-2015POP,Allanson-2016JPP}). However, using standard integral formulae and/or the fact that exponential functions are eigenfunctions of the Weierstrass transform (e.g. see \citet{Wolf-1977}), we pose the following DF as a solution,
\begin{eqnarray}
&&f_s(H_s,p_{xs},p_{ys})=\frac{n_{0s}}{(\sqrt{2\pi}v_{th,s})^3}e^{-\beta_{s}H_{s}}\times\nonumber\\
&&\left(a_{0s}e^{\beta_s(u_{xs}p_{xs}+u_{ys}p_{ys})}+a_{1s}e^{2\beta_s(u_{xs}p_{xs}+u_{ys}p_{ys})}+a_{2s}e^{\beta_s(v_{xs}p_{xs}+v_{ys}p_{ys})}+b_s  \right),\label{eq:DF_ansatz}
\end{eqnarray}
for $a_{0s}, a_{1s}, a_{2s}, b_s, u_{xs}, u_{ys}, v_{xs}$ and $v_{ys}$ as yet arbitrary constants, with the ``$a,b$'' constants dimensionless, and the ``$u,v$'' constants the bulk flows of individual particle populations (e.g. see \citet{Davidsonbook, Schindlerbook}). 

For  the full details describing how the microscopic and macroscopic parameters of the equilibrium are related, and how they are fixed, see the Appendix. In particular, note that $b_{s}$ must satisfy a certain bound in order to guarantee non-negativity of the DF.

\subsection{The distribution function is a sum of four Maxwellians}\label{sec:plots}
The equilibrium DF in equation (\ref{eq:DF_ansatz}) is written as a function of the constants of motion ($H_s, p_{xs},p_{ys}$), which was suitable for constructing an exact equilibrium solution to the Vlasov equation. However, we can write $f_s$ explicitly as a function over phase-space ($z,\boldsymbol{v}$), in a form similar to that in equation (\ref{eq:Maxshift}). The crucial mathematical step is to complete the square in the exponent of equation (\ref{eq:DF_ansatz}) (e.g. see \citet{Schindlerbook}), e.g.
\begin{eqnarray}
e^{-\beta_s(H_s-u_{xs}p_{xs}-u_{ys}p_{ys})}&=&e^{q_s\beta_s(u_{xs}A_x+u_{ys}A_y)}e^{(u_{xs}^2+u_{ys}^2)/(2v_{th,s}^2)}\nonumber\\
&&\times\, e^{-[(v_x-u_{xs})^2+(v_y-u_{ys})^2+v_z^2]/(2v_{th,s}^2)}. \nonumber
\end{eqnarray}
In this manner the DF can be re-written as 
\begin{equation}
f_s(z,\boldsymbol{v})=\frac{1}{(\sqrt{2\pi}v_{th,s})^3}\left[\mathcal{N}_{0s}(z)e^{-\frac{(\boldsymbol{v}-\boldsymbol{V}_{0s})^2}{2v_{th,s}^2}}+\mathcal{N}_{1s}(z)e^{-\frac{(\boldsymbol{v}-\boldsymbol{V}_{1s})^2}{2v_{th,s}^2}}+\mathcal{N}_{2s}(z)e^{-\frac{(\boldsymbol{v}-\boldsymbol{V}_{2s})^2}{2v_{th,s}^2}}+be^{-\frac{\boldsymbol{v}^2}{2v_{th,s}^2}}\right],\label{eq:sumMax}
\end{equation}
for the population density and bulk flow variables (``$\mathcal{N},\boldsymbol{V}$'') defined by
\begin{eqnarray}
\mathcal{N}_{0s}(z)=a_{0}e^{q_s\beta_{s} \boldsymbol{A}\cdot\boldsymbol{V}_{0s}}=a_0e^{-\tilde{z}}\text{sech}\tilde{z},   &\boldsymbol{V}_{0s}=(u_{xs},u_{ys},0), \label{eq:nv1}\\
\mathcal{N}_{1s}(z)=a_{1}e^{q_s\beta_{s} \boldsymbol{A}\cdot\boldsymbol{V}_{1s}}=a_1e^{-2\tilde{z}}\text{sech}^{2}\tilde{z},  &\boldsymbol{V}_{1s}=(2u_{xs},2u_{ys},0),\\
\mathcal{N}_{2s}(z)=a_{2}e^{q_s\beta_{s} \boldsymbol{A}\cdot\boldsymbol{V}_{2s}}=a_2\text{sech}^{2}\tilde{z},  &\boldsymbol{V}_{2s}=(v_{xs},v_{ys},0),\label{eq:nv3}
\end{eqnarray}
and with $a_0, a_1, a_2$ and $b$ defined in the Appendix. It is apparent from consideration of the right-hand side of the definitions of the population densities, that $\boldsymbol{N}_{0s}, \boldsymbol{N}_{1s}$ and $\boldsymbol{N}_{2s}$ are in fact independent of species. Note that $\mathcal{N}_{0s}\to 2a_0$ and $\mathcal{N}_{1s}\to 4a_1$ as $\tilde{z}\to-\infty$; $\mathcal{N}_{0s}\to 0$ and $\mathcal{N}_{1s}\to 0$ as $\tilde{z}\to\infty$; and $\mathcal{N}_{2s}\to 0$ as $\tilde{z}\to \pm \infty$.

The representation of $f_s$ in equation (\ref{eq:sumMax}) has the advantages of having a clear physical interpretation, and of being in a form readily implemented into PIC simulations as initial conditions. Despite the fact that each term of $f_s$ as written in equation (\ref{eq:sumMax}) bears a strong resemblance to $f_{\text{Maxw},s}$ as defined by equation (\ref{eq:Maxshift}), $f_s$ is an exact Vlasov equilibrium DF, whereas $f_{\text{Maxw},s}$ is not. 

\begin{figure}[h!]
 \centering
  \includegraphics[width=24pc]{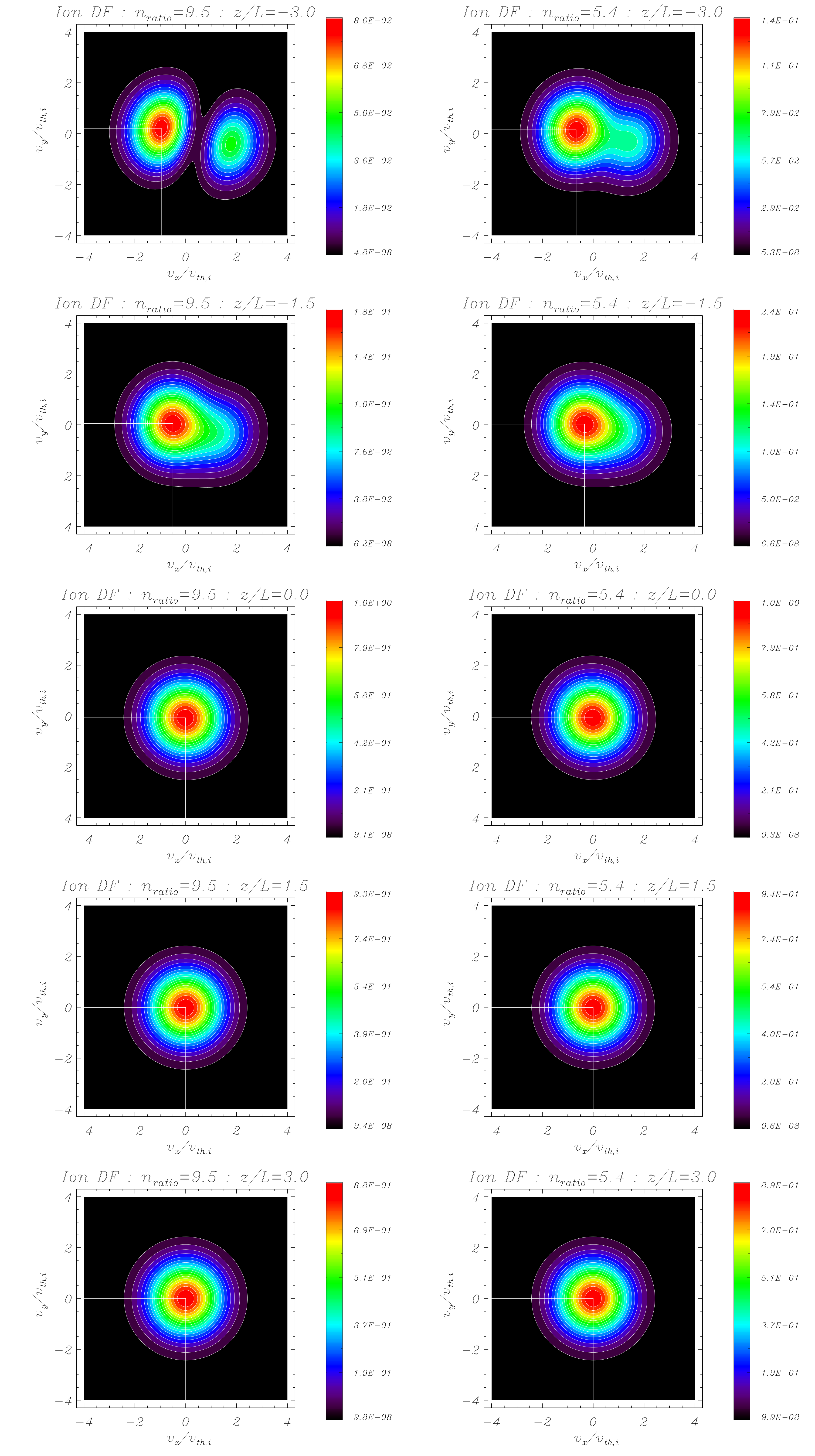}
 \caption{Ion DFs plotted at $\tilde{z}=-3, -1.5, 0, 1.5, 3$, and normalised by $\max_{(v_x,v_y)}f_i(z=0)$. Left-hand column: self-consistent with \emph{Parameter Set One}, i.e. $n_{\text{sheath}}/n_{\text{sphere}}=9.5$. Right-hand column: Self-consistent with \emph{Parameter Set Two}, i.e. $n_{\text{sheath}}/n_{\text{sphere}}=5.4$.}
 \label{figthree}
  \end{figure}

Since the DF is a sum of shifted Maxwellian functions, it is important to understand if, and when, it is possible for the DF to have multiple maxima in velocity space, and/or anisotropies, and how the velocity space structure of the DF depends on the asymmetry of the macroscopic AH+G current sheet equilibrium. A full parameter and/or micro-stability study of the DF is beyond the scope of this paper. However, we show some preliminary results with parameter values that are consistent with asymmetric conditions that could be relevant to PIC modelling of the magnetopause. In Figure \ref{figthree} we plot the ion DF in $(\tilde{v}_x,\tilde{v}_y)$ space, for different $\tilde{z}$ values, and for two sets of parameters. The left-hand column is self-consistent with the macroscopic \emph{Parameter Set One}, whereas the right-hand column is self-consistent with the same magnetic field, but a higher value of $P_T\approx 4.22B_0^2/(2\mu_0)$, such that $n_{\text{sheath}}/n_{\text{sphere}}=5.4$: now known as \emph{Parameter Set Two}. In Figure \ref{figfive} we plot the electron DF for \emph{Parameter Set One} (the electron plots for \emph{Parameter Set Two} are qualitatively very similar). In order to plot the DFs, we must choose values of the constant microscopic parameters that appear in the model. In line with some magnetopause current sheet observations (e.g. \citet{Kaufman-1973, Berchem-1982}), and current PIC approaches (e.g. \citet{Hesse-2013, Liu-2016}), we set the characteristic values of these (constant) microscopic parameters by
\begin{equation}
n_{0i}=1,\hspace{3mm}\delta_i=\frac{m_iv_{th,i}}{eB_0L}=0.1,\hspace{3mm}T_{0i}/T_{0e}=5,\hspace{3mm}\tilde{T}_{0i}+\tilde{T}_{0e}=1.5,\nonumber
\end{equation}
for $\delta_i$ the ratio of the ion thermal Larmor radius to the current sheet width, and $\tilde{T}_{0s}=k_BT_{0s}/(B_0^2/(\mu_0m_in_{0i}))$, i.e. the characteristic temperatures ($k_BT_{0s}=m_sv_{th,s}^2$) are normalised using the characteristic ion Alfv\'{e}n velocity. We also use a realistic mass ratio $m_i/m_e=1836$. The actual values of the plasma magnetisation, temperature, and temperature ratios will of course be position-dependent. Note that both the electron and ion DFs are fully determined once the following parameters are given,
\[
n_{0i},\, \delta_i,\, T_{0i}/T_{0e}, \, \tilde{T}_{0i}+\tilde{T}_{0e},\,m_i/m_e,\, P_T, \, C_1,\, C_2,\, C_3,
\]
and hence the parameter space to investigate is nine-dimensional (in principle one could specify a different set of nine parameters, provided that they are independent).

By contrasting Figures \ref{figthree} and \ref{figfive}, we see immediately that is the ions that carry the `non-Maxwellian' features (anisotropies and possibly multiple peaks) for these parameter values. The ion DFs relevant to \emph{Parameter Set One} seem to suggest that stronger macroscopic asymmetries across the current sheet can be self-consistent with more strongly non-Maxwellian ion DFs. Whereas, those relevant to \emph{Parameter Set Two} demonstrate that it is possible to construct DFs with single maxima in velocity space, whilst still maintaining significant asymmetries across the sheet. However we note that we only present preliminary results here, and a more detailed parameter study will be important to carry out. It may be the case that the ion DFs for \emph{Parameter Set One} are physically unrealistic equilibrium configurations, as they seem susceptible to velocity-space instabilities \citep{Gary-2005} (although the magnitude of the secondary peaks at $\tilde{z}=-3$ are less than $10\%$ of the maximum at $\tilde{z}=0$), whereas those in \emph{Parameter Set Two} may be more realistic. It will be interesting to carry out analytical and/or numerical stability studies in the future.

In Figure \ref{figfive} we plot the ion and electron number densities: $n_s(z,\boldsymbol{v})=\int f_s d^3v$, bulk flows: $\boldsymbol{V}_{s}(z,\boldsymbol{v})=n_s^{-1}\int \boldsymbol{v}f_sd^3v$, and kinetic temperatures: $T_s(z)=(3k_Bn_s)^{-1}(P_{xx}+P_{yy}+P_{zz})$, for \emph{Parameter Set One} (the plots for \emph{Parameter Set Two} are qualitatively similar). The number densities are normalised by the $n_{0s}$ parameter; the $x-$ and $y-$components of the bulk flow are normalised by $|V_{x,0s}+V_{x,1s}+V_{x,2s}|/3$ and $|V_{y,0s}+V_{y,1s}+V_{y,2s}|/3$ respectively; and the temperatures are normalised by the characteristic ion Alfv\'{e}n velocity. These curves demonstrate that it is possible for the DF to be self-consistent with strong density, bulk velocity and kinetix temperature asymmetries across the current sheet. We also see that whilst the DF is not only self-consistent with $j_x=0$, it is also consistent with $\boldsymbol{V}_{xs}=0$, i.e. the independent species bulk flows in the $x$ direction are zero. We also see that the bulk flows in the $y$-direction decay to zero far from the sheet, in contrast to the aforementioned solution put forward by \citet{Alpers-1969}. Hence, our solution has bulk flow properties at the boundaries that are consistent with those of the initial conditions of typical PIC simulations of asymmetric reconnection.

\section{Discussion}
We have presented new, exact and fully self-consistent equilibrium solutions of the Vlasov-Maxwell system in one spatial dimension. Macroscopically, these solutions describe an `asymmetric Harris sheet' magnetic field profile, with finite guide field, such as has often been used in studies of magnetopause current sheets. The expression for the Vlasov equilibrium distribution function is elementary in form, and is written as a sum of four exponential functions of the constants of motion, which can be re-written in $(z,\boldsymbol{v})$ space as a weighted sum of `shifted-Maxwellian' distribution functions.  This form for the distribution function can be readily used as initial conditions in particle-in-cell simulations, and should be particularly suited to studying asymmetric reconnection processes, with potential relevance to. e.g. Earth's magnetopause. The DF is self-consistent with asymmetric profiles of the magnetic field, kinetic temperature, number density and dynamic pressure.

Setting up a current sheet that has an exact Vlasov equilibrium in numerical simulations could be helpful for the study of the collisionless tearing instability, which could be important to understand the nature of intense current sheets at the reconnection x-line. Oblique tearing modes were recently argued to play a potential role in determining the orientation of the three-dimensional reconnection x-line in asymmetric geometry \citep{Liu-2015}, and in causing the bifurcated electron diffusion region in the symmetric geometry \citep{Liu-2013}. The former study is especially crucial for predicting the location of magnetic reconnection at Earth's magnetopause under a diverse solar wind conditions \citep{Komar-2015}. Such an equilibrium solution also facilitates the study of tearing instabilities under the influence of cross-sheet gradients \citep{Zakharov-1992, Kobayashi-2014, Pueschel-2015, Liu-2016}, important to the onset and suppression of sawtooth crashes in fusion devices.

It will be important in the future to further analyse the velocity-space structure of the DF derived in this paper, how it depends on the micro- and macroscopic parameters, and the degree of asymmetry across the current sheet. Also, it will be interesting further work to investigate the practical improvement in a PIC simulation of implementing the DF derived in this paper, as compared to the typical fluid-based equilibrium approach.

\clearpage
\section*{Figures}

  \begin{figure}[h!]
 \centering
 \includegraphics[width=24pc]{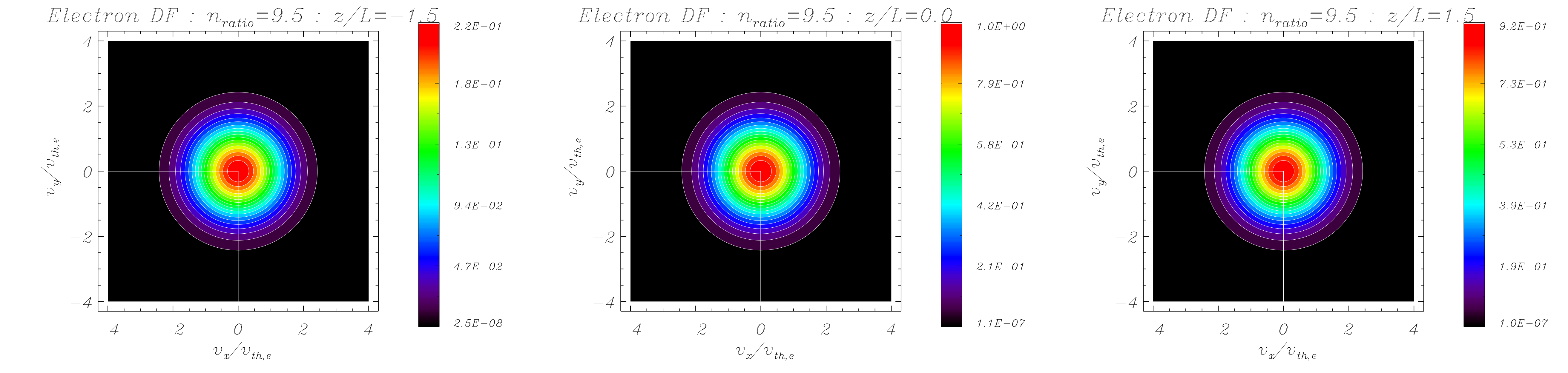}
  \caption{Electron DFs plotted at $\tilde{z}=-1.5, 0, 1.5$, and normalised by $\max_{(v_x,v_y)}f_e(z=0)$. Self-consistent with \emph{Parameter Set One}, i.e. $n_{\text{sheath}}/n_{\text{sphere}}=9.5$.}
 \label{figfive}
   \end{figure}
   
   \begin{figure}[h!]
 \centering
% when using pdflatex, use pdf file:
 \includegraphics[width=25pc]{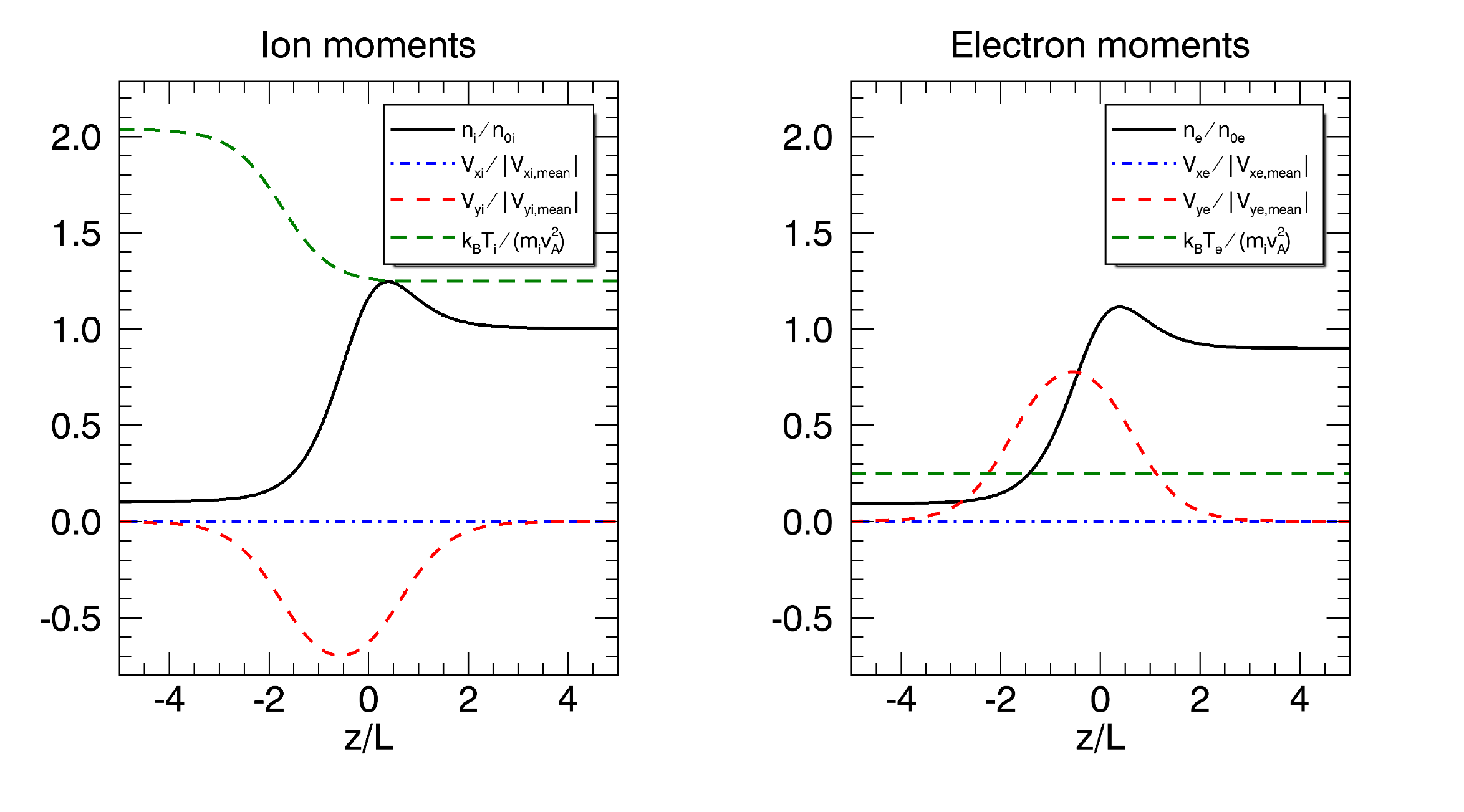}
%
% when using dvips, use .eps file:
% \includegraphics[width=20pc]{figsamp.eps}
%
 \caption{The ion and electron number densities, bulk flows, and temperatures. The number densities, $n_s$, are normalised by the $n_{0s}$ parameter. The components of the bulk flows, $\boldsymbol{V}_{s}$, are normalised by the magnitude of the components of $\boldsymbol{V}_{0s}+\boldsymbol{V}_{1s}+\boldsymbol{V}_{2s}/3$. The temperatures are normalised using the characteristic ion Alfv\'{e}n velocity, $v_{A0}=B_0/\sqrt{\mu_0m_in_{0i}}$. Parameter values: \emph{Parameter Set One}.}
 \label{figfive}
  \end{figure}

\clearpage

\appendix
\section{Equilibrium parameters and their relationships}
We now proceed with the necessary task of ensuring that $n_i(A_x,A_y)=n_e(A_x,A_y)$ (for $n_{s}(A_{x},A_{y})$ the number density of species $s$) in order to be consistent with our assumption that $\phi=0$. The constants $a_0,a_1,a_2$ and $b$ are defined by these neutrality relations, are found by calculating the zeroth order moment of the DF, and are given by
\begin{eqnarray}
a_{0}=n_{0s}a_{0s}e^{(u_{xs}^2+u_{ys}^2)/(2v_{th,s}^2)},  &  a_{2}= n_{0s}a_{2s}e^{(v_{xs}^2+v_{ys}^2)/(2v_{th,s}^2)}, \label{eq:neutral1}\\
 a_{1}=n_{0s}a_{1s}e^{2(u_{xs}^2+u_{ys}^2)/v_{th,s}^2},   &   b=n_{0s}b_s.\label{eq:neutral2}
 \end{eqnarray}
Note that equations (\ref{eq:neutral1}) and (\ref{eq:neutral2}) hold for both ions and electrons ($s=i,e$). We must also ensure that the DF in equation (\ref{eq:DF_ansatz}) exactly reproduces the correct pressure tensor expression of equation (\ref{eq:pressure_tensor}). After some algebra we find the `micro-macroscopic' consistency relations by taking the $v_z^2$ moment of the DF, that complete this final step of the method, and are given by
\begin{eqnarray}
P_{T}-\frac{B_0^2}{2\mu_0}\left[(C_1+C_2)^2+C_3^2\right]=b\frac{\beta_e+\beta_i}{\beta_e\beta_i}, &\displaystyle \frac{C_1-C_2}{C_2C_3B_0L}=e\beta_iu_{xi}=-e\beta_eu_{xe}, \label{eq:micromacro1}\\
4C_1C_2\frac{B_0^2}{2\mu_0}=a_0\frac{\beta_e+\beta_i}{\beta_e\beta_i}, &   \displaystyle\frac{1}{C_2B_0L}=e\beta_iu_{yi}=-e\beta_eu_{ye} ,\label{eq:micromacro2}
 \end{eqnarray}
 \begin{eqnarray}
 -C_1C_2\frac{B_0^2}{2\mu_0}=a_1\frac{\beta_e+\beta_i}{\beta_e\beta_i}, & \displaystyle\frac{2C_1}{C_2C_3B_0L}=e\beta_iv_{xi}=-e\beta_ev_{xe},\\
  C_2(C_2-C_1)\frac{B_0^2}{2\mu_0}=a_2\frac{\beta_e+\beta_i}{\beta_e\beta_i},& \displaystyle\frac{2}{C_2B_0L}=e\beta_iv_{yi}=-e\beta_ev_{ye}. \label{eq:micromacro4}
 \end{eqnarray}

\subsection{Non-negativity of the DF}
 Since we integrate $f_s$ over velocity space to calculate $P_{zz}$, it is clear that non-negativity of $P_{zz}$ does not imply non-negativity of $f_s$. Furthermore, it is clear from equations (\ref{eq:neutral1}) and (\ref{eq:micromacro2}) that $C_1C_2<0$ implies that $a_{0s}<0$ (as well as $a_{1s}>0$, $a_{2s}>0$). By completing the square, the DF can be re-written and we see that non-negativity of the DF is assured provided $b_s\ge a_{0s}^2/(4a_{1s})$.

\acknowledgments
The authors gratefully acknowledge the support of the Science and Technology Facilities Council Consolidated Grant Nos. ST/K000950/1 and ST/N000609/1 (O.A., T.N., J.D.B.H. and F.W.), the Science and Technology Facilities Council Doctoral Training Grant No. ST/K502327/1 (O.A. and J.D.B.H), the Natural Environment Research Council Grant No. NE/P017274/1 (Rad-Sat) (O.A.), the NASA grant NNX16AG75G (Y.-H.L), and NASA's Magnetospheric Multiscale Mission. (Y.-H.L.). O.A. and F.W. would also like to thank the National Science Foundation for support towards attendance at the AGU Chapman Conference on Currents in Geospace and Beyond, 2016. O.A. would also like to thank the Royal Astronomical Society for a travel grant, for support of attendance at the AGU Fall Meeting 2016. 

No significant data sets, models, or modelling techniques have been used by - or newly generated by - the authors for use in this paper. Figures 1, 3, 4 and 5 have all been constructed using standard mathematical functions and basic IDL plotting routines. All of the information that is necessary to reproduce these figures in included in the manuscript.

The authors would like to thank the referees, whose comments have helped to improve the manuscript.

%\bibliography{/Users/OAllanson/Dropbox/biblio}

%%%%%%%%%%%%%%%%%%%%%%%%%%%%%%%%%%%%%%%%%%%%%%%%%%%%%%%%%%%%%%%%%%%%%
% Track Changes:
% To add words, \added{<word added>}
% To delete words, \deleted{<word deleted>}
% To replace words, \replace{<word to be replaced>}{<replacement word>}
% To explain why change was made: \explain{<explanation>} This will put
% a comment into the right margin.

%%%%%%%%%%%%%%%%%%%%%%%%%%%%%%%%%%%%%%%%%%%%%%%%%%%%%%%%%%%%%%%%%%%%%
% At the end of the document, use \listofchanges, which will list the
% changes and the page and line number where the change was made.

% When final version, \listofchanges will not produce anything,
% \added{<word or words>} word will be printed, \deleted{<word or words} will take away the word,
% \replaced{<delete this word>}{<replace with this word>} will print only the replacement word.
%  In the final version, \explain will not print anything.
%%%%%%%%%%%%%%%%%%%%%%%%%%%%%%%%%%%%%%%%%%%%%%%%%%%%%%%%%%%%%%%%%%%%%

%%%
\listofchanges
%%%

\end{document}